\documentclass[a4paper,fleqn]{cas-sc}
\usepackage[numbers]{natbib}
\usepackage[capitalise]{cleveref}
\usepackage{multirow}

\usepackage{array}
\usepackage{threeparttable}
\usepackage{makecell}
\usepackage{fancyvrb}

\DefineVerbatimEnvironment{codeblock}{Verbatim}{fontsize=\small}

\def\tsc#1{\csdef{#1}{\textsc{\lowercase{#1}}\xspace}}
\tsc{WGM}
\tsc{QE}

\begin{document}

\shorttitle{ITKIT}

\shortauthors{Yiqin, et~al.}

\title [mode = title]{ITKIT: Feasible CT Image Analysis based on SimpleITK and MMEngine}

\author[1]{Yiqin Zhang}[orcid=0000-0003-2099-2687]
\cormark[1]
\ead{zyqmgam@163.com}
\ead[url]{https://github.com/MGAMZ}
\credit{Project administration, Conceptualization, Data curation, Formal analysis, Investigation, Methodology, Software, Resources, Validation, Visualization, Writing - Original draft preparation, Writing - review and editing}
\cortext[1]{Corresponding author}

\author[2]{Meiling Chen}
\ead{meiling_chen0313@163.com}
\credit{Formal Analysis, Validation, Visualization, Writing - Reviewing and Editing}

\affiliation[1]{
    organization={Lingang Laboratory},
    city={Shanghai},
    country={China}}
\affiliation[2]{
    organization={Formal-tech},
    city={Shanghai},
    country={China}}

\begin{abstract}
    CT images are widely used in clinical diagnosis and treatment, and their data have formed a de facto standard - DICOM. It is clear and easy to use, and can be efficiently utilized by data-driven analysis methods such as deep learning. In the past decade, many program frameworks for medical image analysis have emerged in the open-source community. ITKIT analyzed the characteristics of these frameworks and hopes to provide a better choice in terms of ease of use and configurability. ITKIT offers a complete pipeline from DICOM to 3D segmentation inference. Its basic practice only includes some essential steps, enabling users with relatively weak computing capabilities to quickly get started using the CLI according to the documentation. For advanced users, the OneDL-MMEngine framework provides a flexible model configuration and deployment entry. This paper conducted 12 typical experiments to verify that ITKIT can meet the needs of most basic scenarios.
\end{abstract}

\begin{graphicalabstract}
\includegraphics[width=\linewidth]{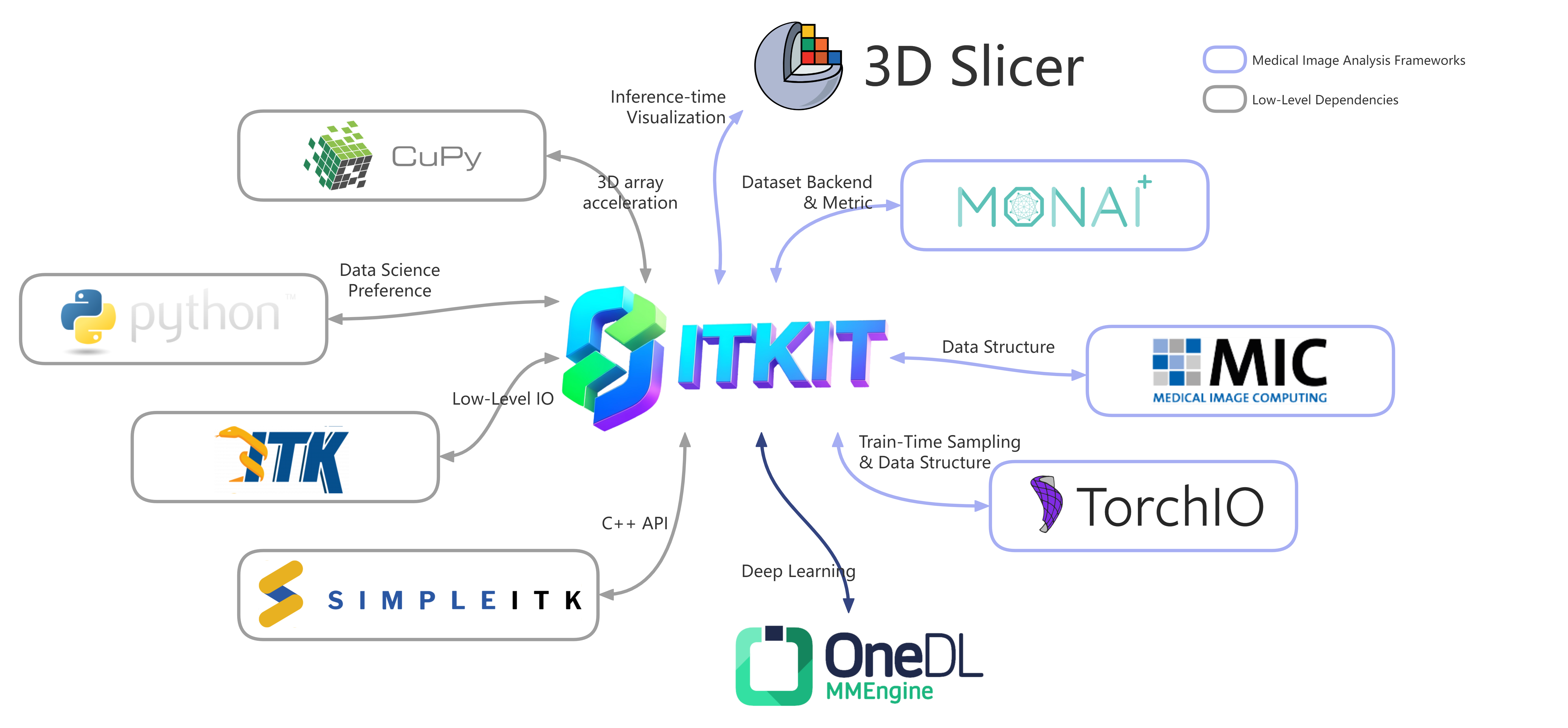}
\includegraphics[width=\linewidth]{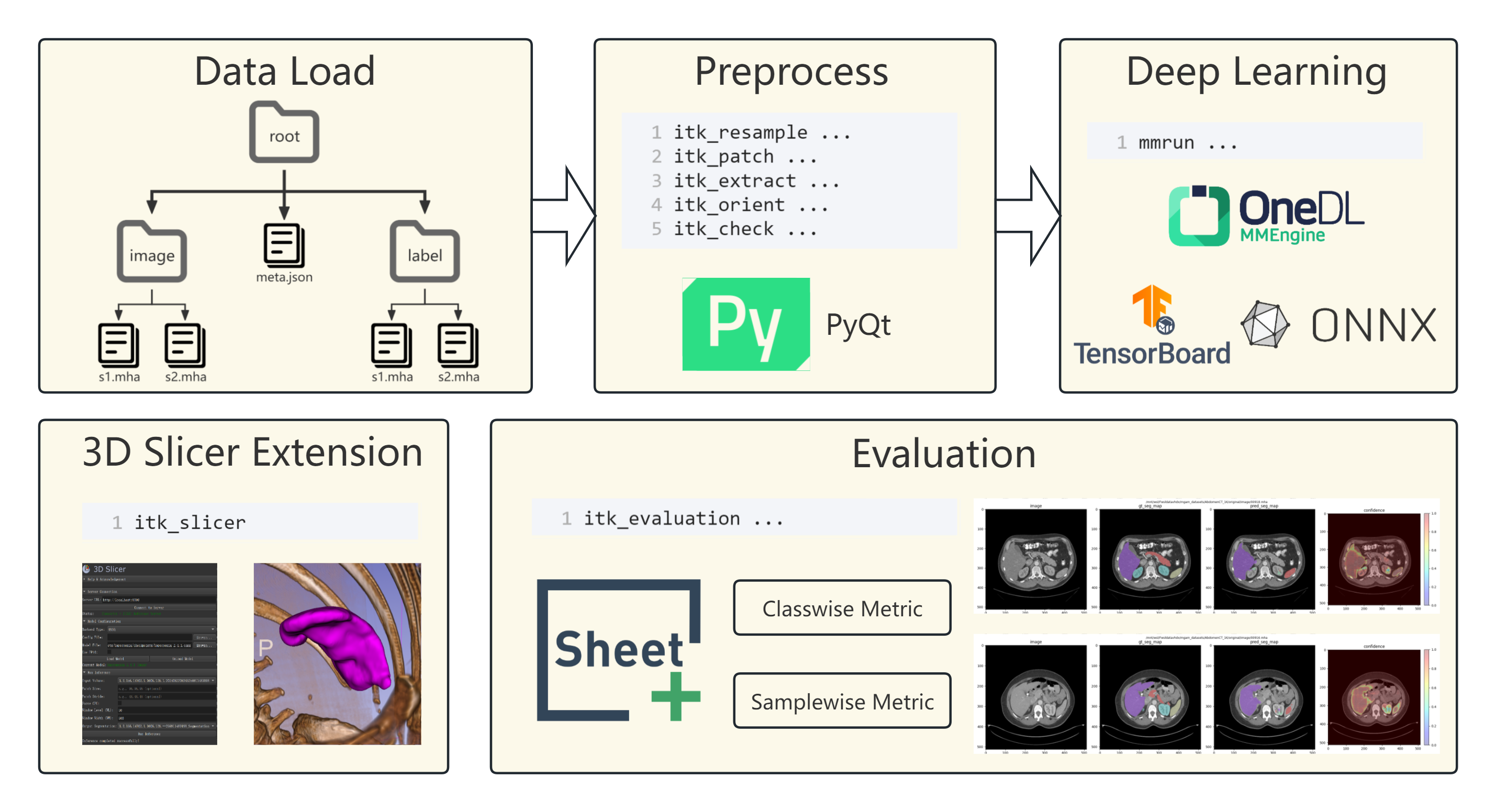}
\includegraphics[width=0.3\linewidth]{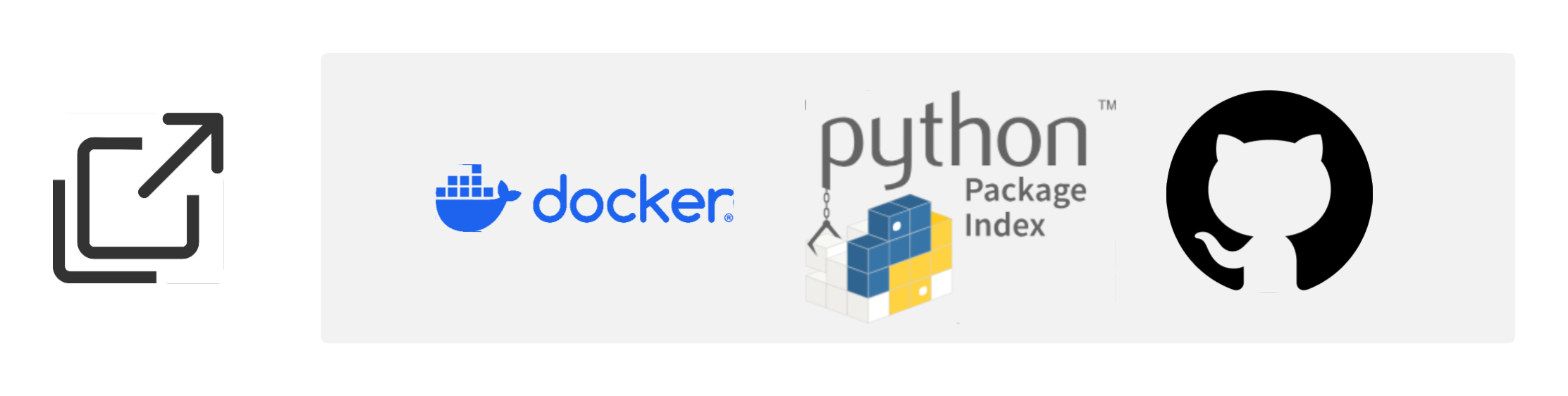}
\end{graphicalabstract}

\begin{highlights}
\item Easy-to-use CT image analysis procedures, covering most fundamental steps.
\item Includes support for a variety of typical datasets and neural networks.
\item The main operations can be achieved through CLI.
\item Work with MONAI, TorchIO, and 3D Slicer.
\end{highlights}


\maketitle

\setlength{\parskip}{\baselineskip}

\section{Introduction}

\subsection{Data-Driven CT Image Segmentation}

Computed Tomography (CT) image segmentation, as one of the core tasks in medical image analysis, plays an irreplaceable role in various clinical scenarios such as organ delineation, lesion quantification, preoperative planning, and treatment response assessment. However, the modeling effectiveness of CT segmentation has long been constrained by the inherent complexity of the data and the fragility of processing pipelines. Prior to the widespread adoption of deep learning, traditional segmentation methods primarily relied on handcrafted features and physics-based prior knowledge, such as thresholding, region growing, active contour models, or graph-cut algorithms. Although these methods offered a certain degree of interpretability, they were highly sensitive to image noise, artifacts, ambiguous tissue boundaries, and variations across imaging devices, making it difficult to achieve robust and consistent performance in real-world clinical settings. More importantly, traditional approaches lacked the capacity to automatically learn common anatomical patterns and pathological variations from large-scale data, severely limiting their generalization ability—particularly when confronted with heterogeneous data collected across multiple centers and imaging protocols.

With the rise of deep learning, medical image segmentation has shifted toward a "data-driven" paradigm centered on end-to-end training. This shift has not only led to significant performance improvements but has also profoundly reshaped researchers' understanding of the fundamental nature of segmentation: it is not merely a pixel-level classification problem, but rather a systems engineering task that critically depends on input data quality, annotation consistency, and spatial semantic integrity.

\subsection{Roles of CT Image Processing Softwares}

Modern medical image segmentation research has gradually converged on a new consensus: the upper bound of model performance is often determined by the quality of data preprocessing. A well-standardized, geometrically aligned, and annotation-validated dataset—even when paired with a relatively simple U-Net variant—can outperform complex Transformer architectures trained on "dirty" data. Consequently, cutting-edge efforts increasingly emphasize building data-centric, engineering-grade pipelines. These pipelines standardize sample structures (e.g., unified organization of image/label/meta files), employ command-line reproducible preprocessing scripts, and incorporate explicit mechanisms to handle common real-world constraints—such as missing labels, non-standard orientations, and extreme voxel spacing. This elevates data preparation from ad-hoc experimental scripts to maintainable, shareable, and verifiable infrastructure. Such a paradigm shift not only enhances the reliability of individual experiments but also enables cross-institutional collaboration, multi-center validation, and long-term model iteration—ultimately bridging the gap between data-driven CT segmentation research and real-world clinical deployment.

In this scenario, a standardized, highly efficient, and low-barrier data preprocessing software represents the optimal solution.

\subsection{Comparison of Other Existing Projects with This Work}

After several years of rapid development, data-driven methodologies and software tools in medical image analysis have become increasingly diverse. Some actively maintained toolkits in the open-source community now enable researchers to rapidly implement most stages—from prototype algorithm validation to clinical deployment. To benchmark our work, several packages with established recognition in the field of medical image analysis were cited and analyzed. As shown in Table~\ref{tab:comparison}, we compared ITKIT with other existing projects from four perspectives: data model, preprocessing, deep-learning, and evaluation. Each perspective was further divided into two subcategories: integrated and command line interface (CLI). An integrated function indicates that the package provides a set of high-level APIs for specific tasks, while a CLI function indicates that the package provides a command line tool for those tasks.

The most widely recognized open-source project is MONAI, supported by NVIDIA. It has a long history, with its components thoroughly tested over time and proven reliable in most scenarios—many libraries depend on it. MONAI covers nearly every aspect of medical image analysis; however, its main repository does not focus heavily on CLI functionality, which is instead implemented through separate sub-repositories. DIPY is another large, high-quality library, notable for providing a comprehensive CLI and offering detailed technical implementations for diffusion, perfusion, and structural imaging in the field of computational anatomy. TorchIO is part of the official PyTorch Ecosystem. It focuses on standardized implementations of common functionalities in this field and integrates well with PyTorch. ITK and its community-driven counterpart, SimpleITK, focus on lower-level aspects such as data formats and I/O design. They are closer to computer hardware and emphasize algorithmic efficiency and standardized data storage practices.

Although these excellent efforts already exist, the research community continues to produce numerous small- to medium-sized libraries that focus on specific needs. These projects typically aim to simplify a particular technical step that is otherwise cumbersome, thereby helping researchers work more efficiently—for example, pymia, PySERA, and Yucca. Many of these libraries originated from real-world challenges encountered by their primary developers during hands-on medical imaging practice.

\begin{table}[tbp]
    \centering
    \caption{Comparison of ITKIT with Other Existing Projects}
    \begin{threeparttable}
        \begin{tabular}{ll|cccccccc}
            \hline
            \multirow{2}{*}{Project} & \multirow{2}{*}{Stars\tnote{a}} & \multicolumn{2}{c}{Data Model} & \multicolumn{2}{c}{Preprocessing} & \multicolumn{2}{c}{Deep-Learning} & \multicolumn{2}{c}{Evaluation} \\
            & & Integrated & CLI & Integrated & CLI & Integrated & CLI & Integrated & CLI \\
            \hline
            MONAI\citep{MONAI} & 7632 & \checkmark & \checkmark & \checkmark & & \checkmark & \checkmark & \checkmark & \\
            ITK\citep{ITK1,ITK2}/S.ITK\citep{SimpleITK} & 1556/1028 & \checkmark & & \checkmark & & & & & \\
            Dipy\citep{Dipy} & 795 & \checkmark & \checkmark & \checkmark & \checkmark & \checkmark & \checkmark & \checkmark & \checkmark \\
            nnUNet\citep{nnU-Net} & 7788 & \checkmark & \checkmark & \checkmark & \checkmark & \checkmark & \checkmark & \checkmark & \checkmark \\
            OpenMedIA & N/A & & & \checkmark & & \checkmark & & \checkmark & \\
            pymia\citep{pymia} & 61 & \checkmark & & \checkmark & & & &  & \\
            PySERA\citep{PySERA} & 17 & \checkmark & & \checkmark & \checkmark & \checkmark & & & \\
            TorchIO\citep{TorchIO} & 2337 & \checkmark & \checkmark & \checkmark & \checkmark & & & & \\
            Yucca\citep{Yucca} & 26 & \checkmark & & \checkmark & \checkmark & \checkmark & \checkmark & \checkmark & \checkmark \\
            p.-m.-t.\citep{Precision-medicine-toolbox} & 70 & & & \checkmark & & & & \checkmark & \\
            \hline
            ITKIT (Ours) & Just Published & \checkmark & \checkmark & \checkmark & \checkmark & \checkmark & \checkmark & \checkmark & \checkmark \\
            \hline
        \end{tabular}
        \begin{tablenotes}
            \item[a] GitHub stars are current as of December 21, 2025.
        \end{tablenotes}
    \end{threeparttable}
    \label{tab:comparison}
\end{table}

For users who are unfamiliar with or dislike coding, a CLI can be a viable option, as it provides a clear entry point and well-defined usage that does not require programming. Aside from DIPY and nnUNet, most projects have not paid significant attention to this aspect. ITKIT aims to optimize precisely in this regard and is especially well-suited for rapid deployment by novice users. It selects canonical methods from all stages of data-driven medical image analysis and provides clear, intuitive CLI. Typically, even clinicians with basic computer literacy can understand the purpose and significance of each CLI command.

With ITKIT, practitioners only need to follow a simple dataset organization convention to effortlessly convert their data into a format that can be efficiently utilized by vision-based neural networks. Moreover, thanks to its built-in efficient neural network architectures, users can quickly obtain a preliminary, usable set of model weights directly through the CLI. Building upon this foundation of rapid prototyping, practitioners can leverage the flexibility of the OpenMM deep learning framework to customize any step of the pipeline—while maintaining strong reproducibility across both data preprocessing and model training workflows.

\section{Methods and Implementations}

\subsection{Data Model}
\label{sec:DataModel}

The ITKIT framework adopts a standardized, pair-centric data model designed to streamline the preprocessing and training pipelines for volumetric medical imaging. At its core, the model enforces a rigid hierarchical directory structure where raw volumetric data and their corresponding semantic annotations are segregated into \textit{image} and \textit{label} subdirectories. This association is maintained through a strict filename-matching protocol, ensuring that each anatomical volume is uniquely coupled with its ground-truth mask.

\begin{figure}
    \centering
    \includegraphics[width=0.5\linewidth]{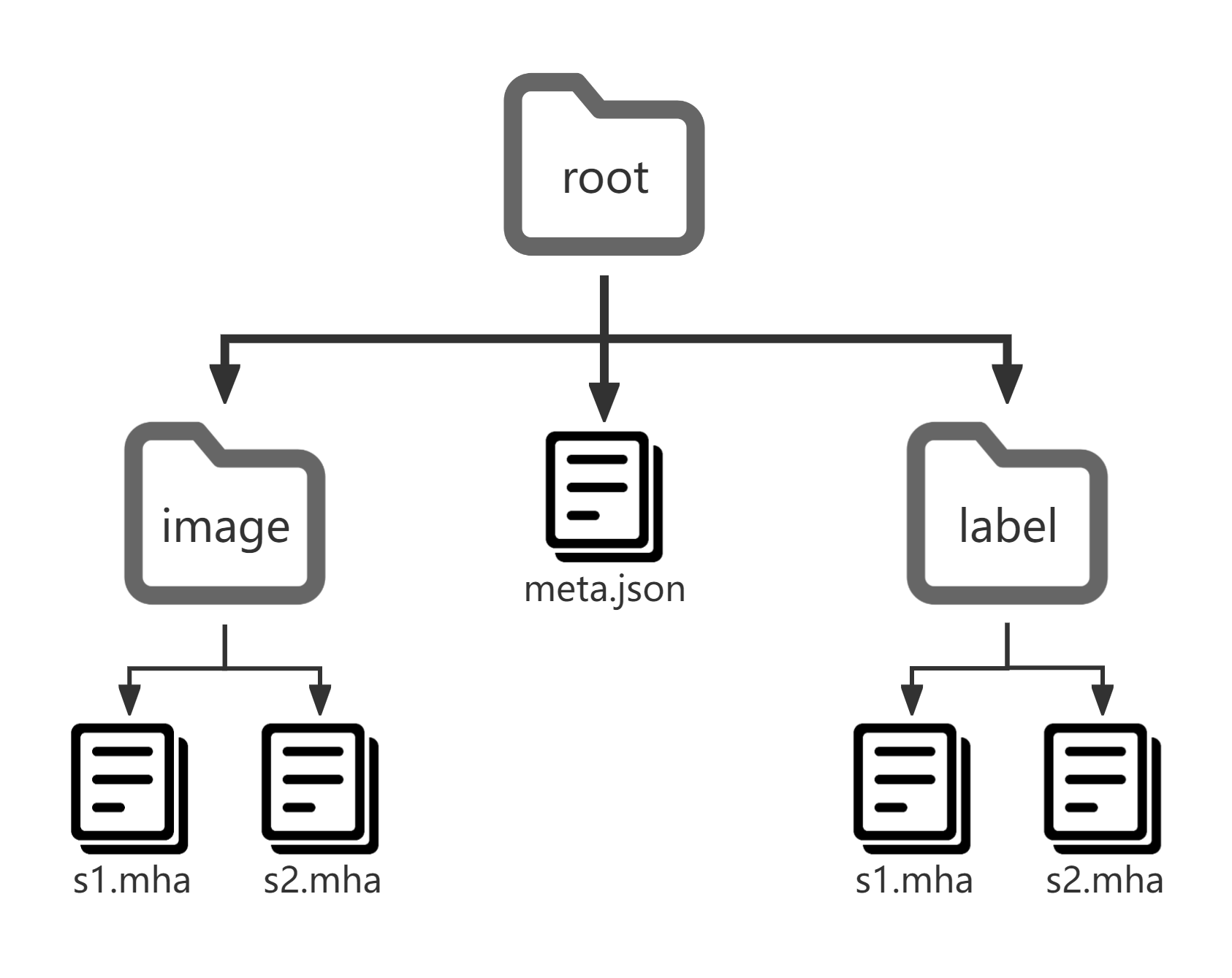}
    \caption{Recommended Dataset Structure for ITKIT. The \texttt{image} and \texttt{label} folders contain the volumetric data and corresponding semantic masks, respectively. The \texttt{meta.json} file stores metadata for the entire dataset, while \texttt{crop\_meta.json} contains metadata for any cropped or patched sub-volumes derived from the original data.}
    \label{fig:Dataset_Structure}
\end{figure}

Central to this architecture is the integration of a global metadata repository, typically encapsulated in a \textit{meta.json} file. This repository serves as a high-level abstraction layer, storing critical geometric attributes such as voxel spacing, volumetric dimensions, and orientation matrices—predominantly following the Left-Posterior-Inferior (LPI) orientation. By decoupling these physical properties from the binary data, the framework enables efficient dataset-wide filtering and validation—such as minimum size or spacing constraints—without the computational overhead of file I/O.

Furthermore, the data model extends its utility to handle derived data structures, such as pre-computed patches and class-specific extractions. Through auxiliary metadata files like \textit{crop\_meta.json}, the framework tracks the provenance and spatial context of sub-volumes and remapped labels. This multi-tiered metadata approach, managed by ITKIT tools, ensures that even complex, multi-modal datasets remain consistent and accessible. By providing built-in conversion CLI to MONAI and TorchIO, the ITKIT data model serves as a versatile bridge between raw medical data and modern deep learning frameworks.

\subsection{Processing Toolchain}

\begin{figure}[pos=tbp]
    \centering
    \includegraphics[width=\linewidth]{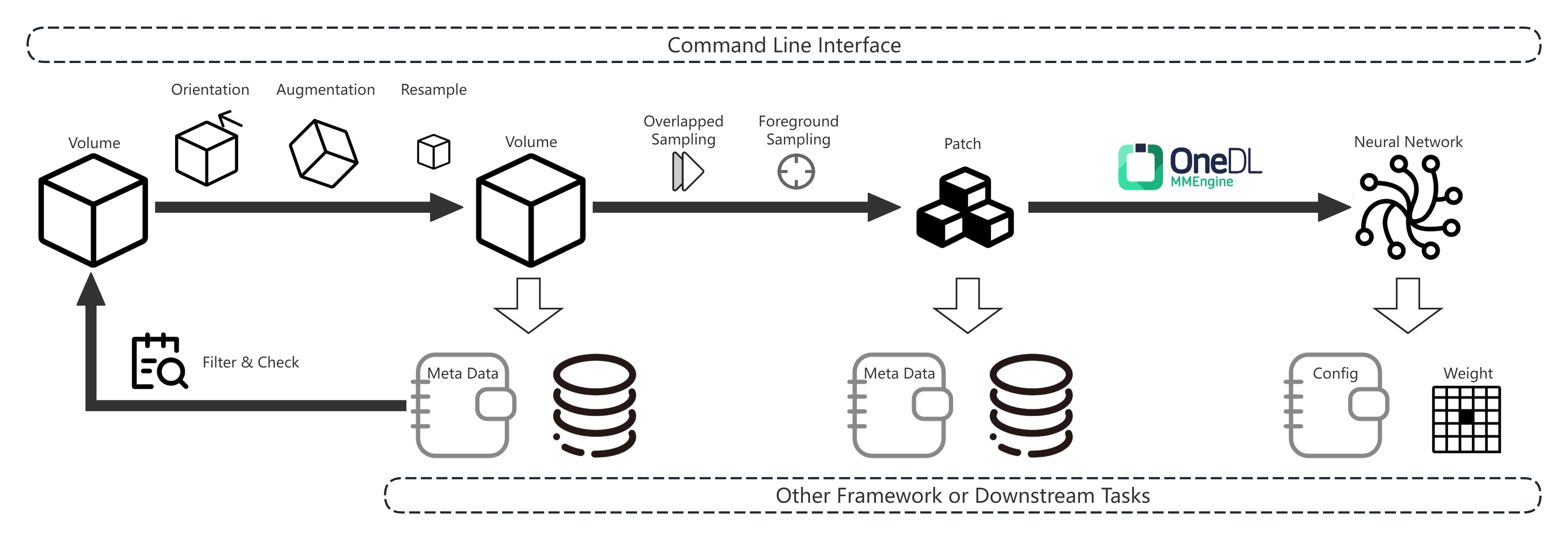}
    \caption{Scope of ITKIT's Functional Abstractions}
    \label{fig:ITKIT_Functions}
\end{figure}

ITKIT provides a concise and clear CLI for several fundamental steps in medical image analysis. These include: dataset metadata inspection, orientation adjustment, resampling, patch splitting, data augmentation, partial label extraction and label remapping, dataset structure conversion, and semantic segmentation accuracy evaluation. Additionally, the mmrun command can be used to invoke ITKIT's neural network training framework, which is built upon OneDL-MMEngine. In other words, researchers can use ITKIT to complete the entire workflow—from data ingestion all the way through neural network training—while also being able to easily select and invoke specific functionalities via the CLI, as is illustrated in \cref{fig:ITKIT_Functions}.

Any of the commands described above is compatible with the Data Model outlined in \cref{sec:DataModel}. Whenever modifications are made to the data itself, ITKIT automatically updates the associated metadata table to maintain consistency between the metadata and the data. Users can invoke the CLI multiple times independently, progressively transforming a dataset into the desired state while being able to inspect the data at each stage using various tools (e.g., ITK-SNAP, 3D Slicer). For certain sampling methods, such as resampling and patch splitting, a processing configuration file is generated alongside the output data, enabling future users to clearly identify the exact procedure used to generate the data.

\subsection{OneDL-MMEngine-Based Deep Learning}

\subsubsection{Task Definition}

ITKIT emphasizes usability and versatility. For data-driven downstream tasks, it implements the most representative 3D medical image segmentation task, supporting both 2D and 3D modes—the latter being more common for CT and MRI modalities. Like most other frameworks, we implement sliding-window inference. In 3D mode, ITKIT leverages the latest PyTorch and CUDA frameworks, employing asynchronous Host-to-Device and Device-to-Host data streaming to accelerate processing of large volumes on PC hardware.

In recent years, the medical image analysis community has largely focused on proposing new neural network architectures and improving accuracy. Against this backdrop, most researchers typically modify only neural network-related configurations while keeping the higher-level task pipeline unchanged. The task design currently provided by ITKIT is particularly well-suited for this scenario, as it allows researchers to focus solely on modifying the neural network without having to implement other data processing components from scratch.

ITKIT leverages OneDL-MMEngine for neural network framework extension—an flexible, configuration-driven framework that ensures strong reproducibility. In fact, several issues identified and improvements made within ITKIT have already been merged back into OneDL-MMEngine. The close collaboration between the two projects helps maintain well-decoupled functionalities, enabling ITKIT users to focus on medical imaging research while confidently relying on a relatively stable underlying training framework.

\subsubsection{Integrated Datasets}

As the starting point of data-driven medical image analysis, data is undeniably foundational. Generally speaking, methods validated on public datasets enjoy higher credibility, while studies conducted on private datasets can provide richer clinical metadata. ITKIT supports representative public datasets from recent years. Although the project does not redistribute any data, once users download a dataset locally in its entirety, they can quickly launch experiments with minimal configuration using ITKIT. The validated and supported datasets are listed in \cref{tab:datasets}. CT or MRI Datasets not listed is likely to be supported with a few configurations.

\begin{table}[tbp]
    \raggedright
    \caption{Supported Public Datasets in ITKIT.}
    \begin{threeparttable}
        \begin{tabular}{lll}
            \hline
            Dataset Name & Number of Classes\tnote{a} & Modality \\
            \hline
            AbdomenCT-1K\citep{AbdomenCT-1K} & 5 & CT \\
            CT-ORG\citep{CT-ORG} & 6 & CT \\
            CT-Spine1K\citep{CTSpine1K} & 26 & CT \\
            FLARE 2022\citep{FLARE22} & 14 & CT \\
            FLARE 2023\citep{FLARE23} & 15 & CT \\
            Image-TBAD\citep{ImageTBAD} & 4 & CT \\
            KiTS23\citep{KiTS19,KiTS21} & 4 & CT \\
            LiTS\citep{LiTS} & 3 & CT \\
            LUNA16\citep{LUNA16} & 2 & CT \\
            Totalsegmentator\citep{TotalSegmentator} & 118 & CT \\
            \hline
        \end{tabular}
        \begin{tablenotes}
            \item[a] The number of classes includes the background class.
        \end{tablenotes}
    \end{threeparttable}
    \label{tab:datasets}
\end{table}

\subsubsection{Integrated Neural Networks}

In recent years, neural networks have become the dominant algorithmic approach in data-driven medical image analysis. Drawing on medical imaging practices from the past few years, ITKIT has selected several classic neural network architectures based on CNNs, Transformers, and Mamba, which researchers can easily deploy and use. In most cases, these models can serve as competitive baselines. Detailed information about these neural networks is provided in \cref{tab:models}.

\begin{table}[tbp]
    \raggedright
    \caption{Integrated Neural Networks in ITKIT.}
    \begin{threeparttable}
        \begin{tabular}{lllll}
            \hline
            Model & Year\tnote{a} & Architecture & Dimension\tnote{b} & Install \& Run \\
            \hline
            UNet 3+\citep{UNet3Plus} & 2020 & CNN & 2D/3D & \checkmark \\
            EfficientNet\citep{EfficientNet} & 2021 & CNN & 2D & \checkmark \\
            SegFormer\citep{SegFormer} & 2021 & Transformer & 2D & Requires Official Repo \\
            DSNet\citep{DSNet} & 2021 & CNN & 2D & \checkmark \\
            UNETR\citep{UNETR} & 2022 & Transformer & 3D & \checkmark \\
            EfficientFormer\citep{EfficientFormer} & 2022 & Transformer & 2D & \checkmark \\
            DconnNet\citep{DconnNet} & 2023 & CNN & 2D & \checkmark \\
            EGE-UNet\citep{EGE-UNet} & 2023 & CNN & 2D & \checkmark \\
            MedNeXt\citep{MedNeXt} & 2023 & CNN & 3D & \checkmark \\
            DA-TransUNet\citep{DA-TransUNet} & 2024 & Transformer & 2D & \checkmark \\
            LM-Net\citep{LM-Net} & 2024 & CNN & 2D & Requires natten \\
            SegFormer3D\citep{SegFormer3D} & 2024 & Transformer & 3D & \checkmark \\
            Swin-UMamba\citep{Swin-UMamba} & 2024 & Mamba & 3D & Requires mamba\_ssm \\
            SegMamba\citep{SegMamba} & 2024 & Mamba & 3D & Requires mamba\_ssm \\
            VMamba\citep{VMamba} & 2024 & Mamba & 3D & Requires mamba\_ssm \\
            \hline
        \end{tabular}
        \begin{tablenotes}
            \item[a] The date when the model's paper was first published.
            \item[b] The dimensionality in which the model performs feature extraction. This does not necessarily indicate the required input data dimensionality.
        \end{tablenotes}
    \end{threeparttable}
    \label{tab:models}
\end{table}

\subsubsection{Integrated Pipelines}

ITKIT employs a data processing pipeline based on OneDL-MMCV and has implemented several commonly used processing steps.

For the loading process, it supports formats such as NIfTI, DICOM, and MetaImage via SimpleITK, noted ITKIT's best practices recommend using the MHA format. In the preprocessing stage, it provides standardization methods including basic window-level control and instance normalization. For data augmentation, it offers a variety of methods such as RandomRoll, RandomFlip, RandomContinuousErase, RandomDiscreteErase, RandomRotate3D, and RandomGamma.

\subsection{Inferencer and 3D Slicer Extension}

ITKIT supports multiple inference methods. The MMEngine backend enables rapid deployment and inference, allowing execution directly from configuration files and weights upon training completion. The ONNX inference backend requires converting weights into the ONNX model format, offering superior cross-model and cross-environment compatibility.

\begin{figure}[pos=tbp]
    \centering
    \includegraphics[width=\linewidth]{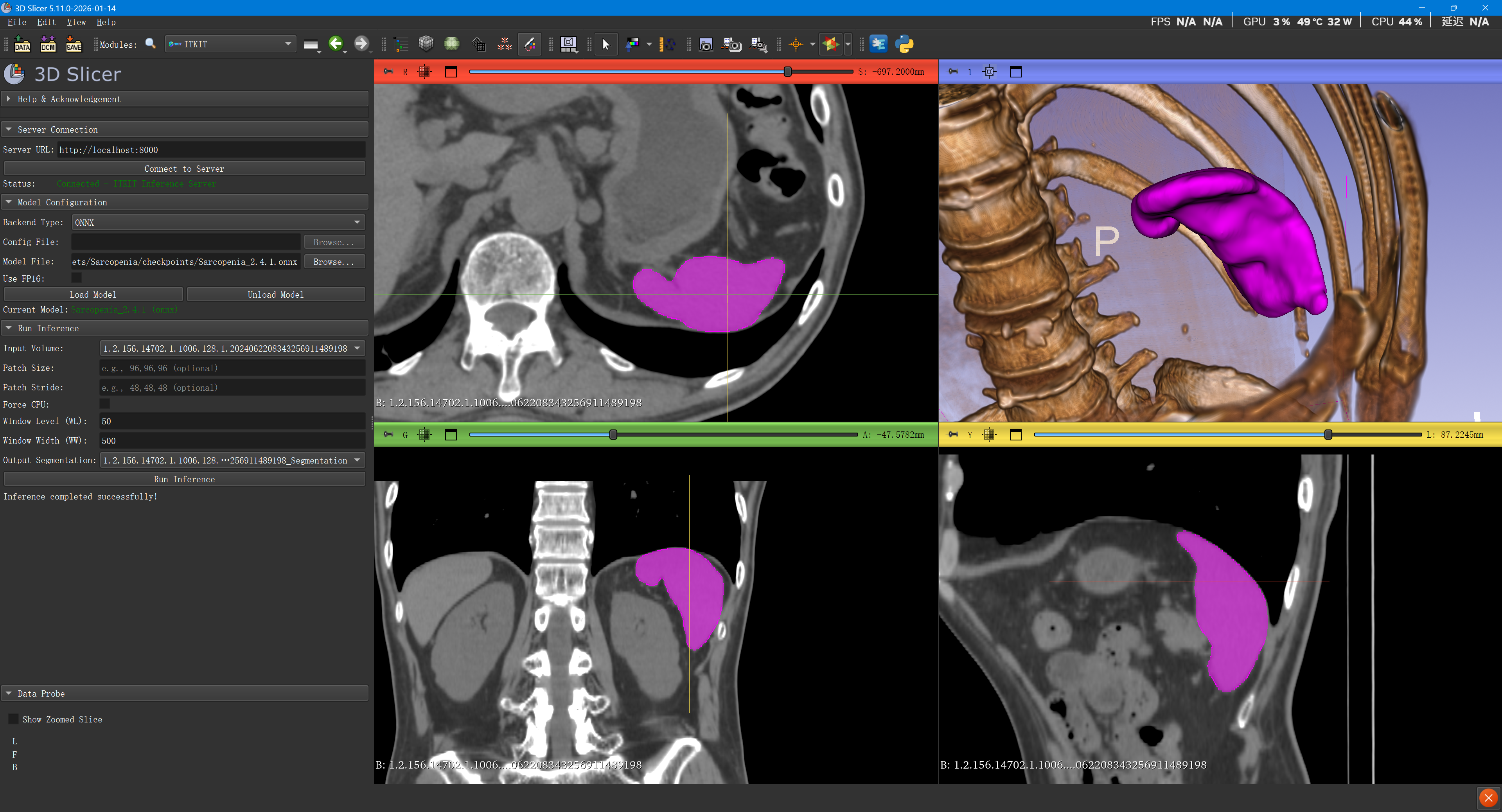}
    \caption{ITKIT 3D Slicer Extension Screenshot. The left side displays the ITKIT extension, providing an interface for configuring the parameters required to perform segmentation inference. The right side shows the results sematic mask of spleen, generated using an ONNX model trained and converted by ITKIT. In this case, the segment engine runs in a Docker container under WSL on Windows.}
    \label{fig:ITKIT_3DSlicer_MainScreenShot}

    \includegraphics[width=\linewidth]{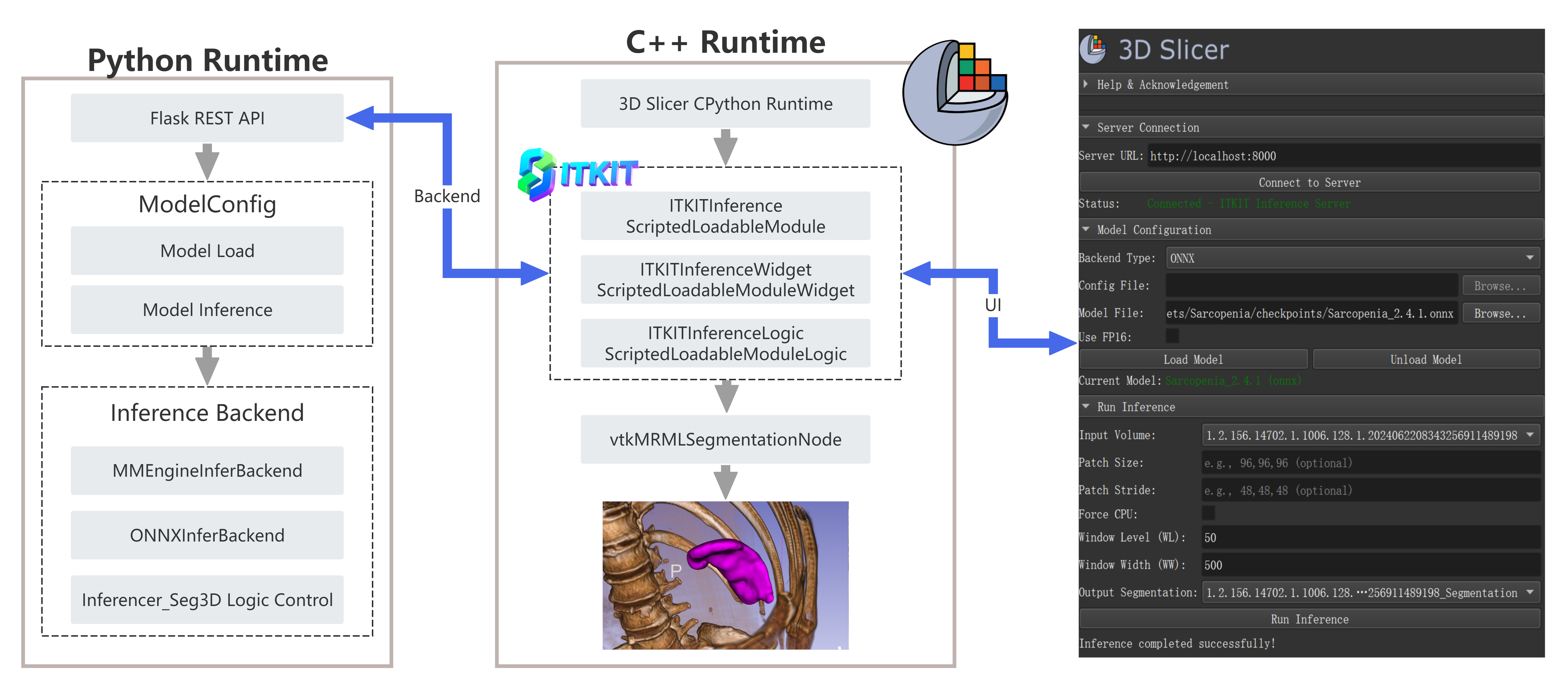}
    \caption{Architecture of ITKIT 3D Slicer Extension. The frontend is built using the QT framework integrated into 3D Slicer, while the backend is reached through Flask. This decoupled design allows the inference engine to run in isolated environments such as Docker containers, WSL, or any remote servers.}
    \label{fig:ITKIT_3DSlicer_Architecture}
\end{figure}

ITKIT provides a 3D Slicer \citep{3D_Slicer} extension to enable real-time inference and visualization(\cref{fig:ITKIT_3DSlicer_MainScreenShot}). This extension supports ONNX or native ITKIT-OneDL-mmengine models and allows the specification of common inference parameters. Like MONAILabel\citep{MONAILabel1,MONAILabel2}, it employs a decoupled frontend-backend architecture(\cref{fig:ITKIT_3DSlicer_Architecture}): the frontend utilizes the QT framework built into 3D Slicer to define the user interface, while the extension communicates with a backend inference engine via Flask upon receiving a request. All neural network inference tasks can be executed in an isolated environment. This design offers deployment flexibility. Users can install heavy dependencies such as PyTorch and CUDA within Docker, WSL, or even on remote servers, requiring only the 3D Slicer software itself to be installed locally.

\section{Results}

To evaluate the generalization capability of ITKIT, we conducted end-to-end segmentation training on AbdomenCT1K tasks using three different data loading backends: the native backend, TorchIO, and MONAI. The neural networks employed during training were MedNeXt and SegFormer3D, representing convolutional and Transformer architectures, respectively. A total of 12 experiments were performed, and their results are summarized in \cref{tab:segment_results}.

\begin{table}[tbp]
    \raggedright
    \caption{Baseline Segmentation Results.}
    \begin{threeparttable}
        \begin{tabular}{lll|lllll}
            \hline
            Task & Data Backend & Model & Duration(s) & Dice & IoU & Recall & Precision \\
            \hline
            \multirow{6}{*}{AbdomenCT1K} & \multirow{2}{*}{Native} & MedNeXt & 119039 & 83.41 & 75.98 & 83.43 & 83.69 \\
            & & SegFormer3D & 17493 & 87.25 & 79.84 & 84.53 & 90.31 \\
            & \multirow{2}{*}{TorchIO} & MedNeXt & 135806 & 88.25 & 80.88 & 85.94 & 91.36 \\
            & & SegFormer3D & 25226 & 94.08 & 89.50 & 94.91 & 93.29 \\
            & \multirow{2}{*}{MONAI} & MedNeXt & 84437 & 88.66 & 81.69 & 85.92 & 92.41 \\
            & & SegFormer3D & 24037 & 89.88 & 83.27 & 86.76 & 93.54 \\
            \hline
        \end{tabular}
        % \begin{tablenotes}
            
        % \end{tablenotes}
    \end{threeparttable}
    \label{tab:segment_results}
\end{table}

\begin{figure}
    \centering
    \includegraphics[width=\linewidth]{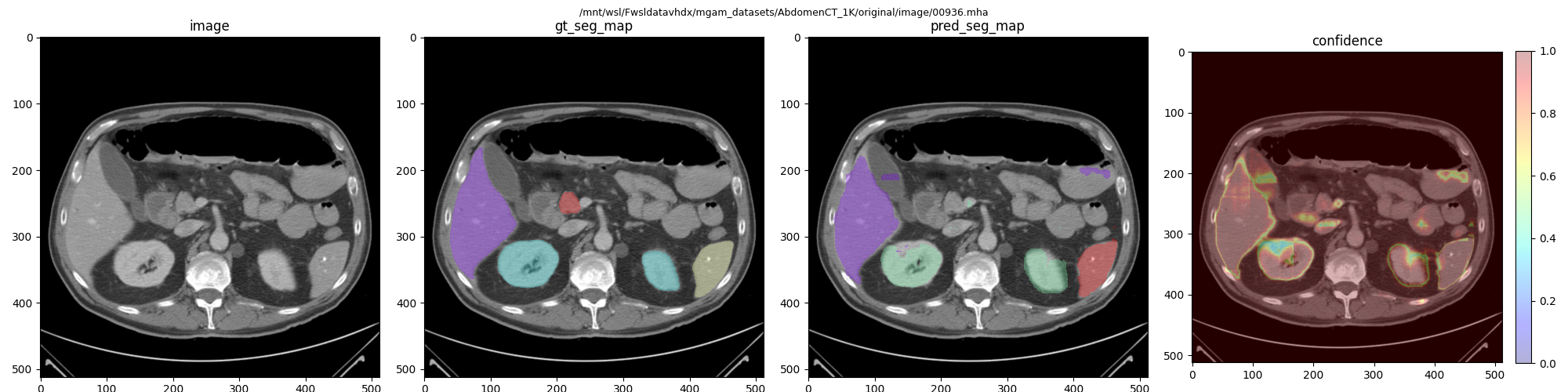}
    \caption{ITKIT Visualization. The image is saved to both local files and TensorBoard during training.}
    \label{fig:ITKIT_TensorBoard}
\end{figure}

During training and the testing phase afterward, ITKIT uses the OneDL-MMEngine framework to provide a custom Visualizer. This tool can display basic CT image slices and record training parameters, including the original images, ground truth, confidence, loss, metrics, etc. All these data are saved both to local files and to TensorBoard, making it easy for users to analyze. \cref{fig:ITKIT_TensorBoard} shows an example of ITKIT visualization during training.

\section{Discussion}

ITKIT is distilled from several previous medical image analysis studies\citep{PSBPD,AutoWindow,SliceWiseAug,SpatialAwareSelfSup} and integrates most of the software processes required for CT image segmentation. As a new open-source software, it is bound to have many bugs. Actively maintaining it and communicating with users will be our focus in the coming period of time.

From a functional perspective, after the community has experienced the frequent development of neural networks and the sharing of datasets in the past few years, the frequency of new SOTA methods and high quality dataset appearing seems to be decreasing. ITKIT hopes to keep up with and support novel and widely recognized neural network architectures and datasets in practice. This ensures that anyone who starts using this software at any time can keep up with the most advanced methods in this field.

\section*{Ethics statement}

The conduct of this research did not interfere with or influence any clinical diagnostic or treatment processes. All the imaging data is anonymous and publicly available.

\section*{Declaration of competing interest}

The authors declare that they have no known competing financial interests or personal relationships that could have appeared to influence the work reported in this paper.

\section*{Data availability}

Users and researchers can access ITKIT at multiple locations.

GitHub Repository: \url{https://github.com/MGAMZ/ITKIT}

PyPI Package: \url{https://pypi.org/project/itkit}

TestPyPI for pre-releases: \url{https://test.pypi.org/project/itkit}

Docker Hub Image: \url{https://hub.docker.com/repository/docker/mgam1009/itkit}

\section*{Declaration of generative AI in scientific writing}

Generative AI and AI-assisted technologies are only used in the writing process to improve the readability and language of the manuscript. The author team hereby affirms that AI was not utilized in any capacity beyond the aforementioned aspects, and assumes full responsibility for all content and expressions presented in this article.

\printcredits

\bibliographystyle{cas-model2-names}
\bibliography{ITKIT.bib}

\clearpage

\appendix

\section{Steps to Reproduce the Baseline}

Here we assume the \texttt{AbdomenCT1K} and \texttt{python>=3.10} environment are already prepared. Users can install ITKIT simply with pip, but for comprehensive reproduce, we recommend clone the repository and install with \texttt{advanced} and \texttt{onnx} extra dependencies, as the repository also contains example configs and readmes. According to ITKIT best practices, the datasets should be organized in the file structure as \cref{fig:Dataset_Structure}.

Users can use \texttt{itk\_resample} to normalize the voxel spacing between samples and \texttt{itk\_check symlink} to filter out samples that do not meet the minimum size requirements. For \texttt{TorchIO} and \texttt{MONAI} backends, data preprocessing is completed at this point. For the native backend, users need to use \texttt{itk\_patch} to patch the 3D volumes into smaller 3D sub-volumes, which can provide lower CPU overhead.

\begin{codeblock}
git clone https://github.com/MGAMZ/ITKIT
pip install "./ITKIT[advanced,onnx]"
itk_resample \
    dataset \
    /path/to/source_dataset \
    /path/to/resampled_dataset \
    --spacing 2 2 2 \ # [Z, Y, X]
    --mp
itk_patch \
    /path/to/resampled_dataset \
    /path/to/patched_dataset \
    --patch-size 96 96 96 \ # [Z, Y, X]
    --patch-stride 48 48 48 \ # [Z, Y, X]
    --mp
\end{codeblock}

All training configuration follows \texttt{OneDL-MMEngine} specification. Users have to modify the dataset paths in the config files prepared at \texttt{./ITKIT/examples}. Noted that \texttt{MONAI} and \texttt{TorchIO} backends only require one dataset path, while the native backend requires two paths for original volumes and patched volumes. After specifing the correct config file, users have to set several environment variables to setup the working directories as follows:

\begin{codeblock}
export mm_workdir="/path/to/work_dir"
export mm_testdir="/path/to/test_dir"
export mm_configdir="/path/to/ITKIT/examples"
\end{codeblock}

With all the preparation done, users can start training with \texttt{mmrun} command to queue the experiments. Below is an example command to run example experiments. Each version number corresponds to a config folder, and each folder contains two neural networks which will be detected by \texttt{mmrun} automatically. So the following command will launch 6 experiments in total.

\begin{codeblock}
mmrun "0.0" "0.1" "0.2"
\end{codeblock}

\end{document}